\documentclass[namedreferences]{SolarPhysics}
\usepackage[optionalrh]{spr-sola-addons} 
\usepackage{graphicx}        
\usepackage{color}           
\usepackage{url}             

\begin{document}
\begin{article}
\begin{opening}

\title{Diagnosing the Source Region of a Solar Burst on 26 September 2011 by Microwave Type III Pairs}

\author{Baolin~\surname{Tan}$^{1, 2}$} \runningauthor{Tan et al.}
\author{Marian~\surname{Karlick\'y}$^{3}$}
\author{Hana~\surname{M\'esz\'arosov\'a}$^{3}$}
\author{Larisa~\surname{Kashapova}$^{4}$}
\author{Jing~\surname{Huang}$^{1, 2}$}
\author{Yan~\surname{Yan}$^{1, 2}$}
\author{Eduard P.~\surname{Kontar}$^{5}$}
   \institute{$^{1}$ National Astronomical Observatories, Chinese Academy of
Sciences, Beijing 100012, China, Email: \url{bltan@nao.cas.cn}\\
$^{2}$ School of Astronomy and Space Sciences, University of Chinese Academy of Sciences, Beijing 100049, China\\
$^{3}$ Astronomical Institute of the Academy of Sciences of the
Czech Republic, CZ--25165 Ond\v{r}ejov, Czech Republic\\ $^{4}$
Institute of Solar-Terrestrial SB RAS, Lermontov st. 126a,
6640333, Irkutsk, Russia\\ $^{5}$School of Physics \& Astronomy,
University of Glasgow, G12 8QQ, Glasgow, Scotland, UK
\author{Hana~\surname{M\'esz\'arosov\'a}$^{2}$\sep}
 \runningtitle{Diagnosing the Source Region by Microwave Type III Burst Pairs}\\
             }

\begin{abstract}

This work reports a peculiar and interesting train of microwave
type III pair bursts in the impulsive rising phase of a solar
flare on 2011 September 26. The observations include radio
spectrometers at frequency of 0.80 - 2.00 GHz, hard X-ray (RHESSI
and FERMI), EUV images of SWAP/PROBA-2 and  magnetogram of
HMI/SDO. By using a recently developed method (Tan et al. 2016a),
we diagnosed the plasma density, temperature, plasma beta,
magnetic field near the source region, the energy of energetic
electrons and the distance between the acceleration region and the
emission start sites of type III bursts. From the diagnostics, we
find that: (1) The plasma density, temperature, magnetic field,
and the distance between the acceleration region and the emission
start sites almost have no obvious variations during the period of
type III pair trains, while the energy of electrons has an obvious
peak value which is consistent to the hard X-ray emission. (2) The
plasma beta is much higher than an unity showing a highly dynamic
process near the emission start site of type III bursts. (3)
Although the reversed-slope type III branches drift slower at one
order of magnitude than that of the normal type III branches, the
related downgoing and upgoing electrons still could have same
order of magnitude of energy. These facts indicate that both of
the upgoing and downgoing electrons are possibly accelerated by
similar mechanism and in a small source region. This diagnostics
can help us to understand the microphysics in the source region of
solar bursts.

\end{abstract}
\keywords{Sun: microwave emission --- Sun: magnetic reconnection
--- Sun: flares}
\end{opening}

\section{Introduction}
     \label{S-Introduction}

Accelerated electron beams are believed to be responsible for both
hard X-ray (HXR) and strong coherent radio emission during solar
flares. However, so far the location of the electron acceleration
and its physical parameters are poorly known. Solar microwave type
III pair burst is possibly the most sensitive signature of the
primary energy release and electron accelerations in flares
(Aschwanden et al. 1997). A type III pair is composed of two type
III burst branches beginning almost at the same time, one is a
normal type III burst with negative frequency drift and the other
is a reverse-sloped (RS) type III burst with positive frequency
drift. The normal branch can track the upward energetic electron
beam, and the RS branch is associated to the downward electron
beam (Robinson \& Benz 2000). In an ideal condition, a type III
pair is made up of a normal branch and a RS branch simultaneously
(Aschwanden et al. 1993, Huang et al. 1998, Ning et al. 2000).
Practically, observations show a group of normal type III bursts
at low frequency band and a group of RS type III bursts at higher
frequency band occur in same time interval (Aschwanden et al.
1997, Ma et al. 2008, Tan et al. 2016b) which might reflect the
complex irregular magnetic structures, repeatedly electron
acceleration, and rapid changes in the source regions (Benz et al.
1992, Meshalkina et al. 2004). It is difficult to clarify their
one-to-one corresponding relationships. We call this complex
assembly a type III pair train.

Microwave type III pairs can be regarded as a sensitive tool to
diagnose the physical conditions around the source region of solar
bursts where magnetic reconnection, energy release, and particle
acceleration take place (Aschwanden \& Benz 1997, Sakai et al.
2005, Altyntsev et al. 2007, M\'esz\'arosov\'a et al. 2008, Li et
al. 2011, Reid, Vilmer, \& Kontar, 2011). This work reports a
well-observed microwave type III pair train in a solar HXR bursts
on 2011 September 26. It is so peculiar that we have never seen so
clear microwave type III pair train in impulsive rising phase of
the previous solar flares. We apply a recent developed method (Tan
et al. 2016a) to derive the physical parameters near the start
sites of the above type III bursts. Section 2 presents the
observing properties of the microwave type III pair train and the
related hard X-ray emission and source information. Section 3
presents the diagnostic results. And finally the conclusion and
discussions are summarized in Section 4.

\section{Observation of the microwave type III pair train}
      \label{S-general}

\subsection{Observation of Radio Spectrometer}

The microwave type III pair train is observed by the Ond\v{r}ejov
radiospectrograph in the Czech Republic (ORSC). ORSC is an
advanced dynamic spectrometer which locates at Ond\v{r}ejov, the
Czech Republic. It receives the solar radio total flux at
frequencies of 0.80 - 5.00\,GHz with cadence of $\triangle t=10$
ms and frequency resolution of $\triangle f=5$ MHz at frequency of
0.80 - 2.00\,GHz and $\triangle f=12$ MHz at frequency of 2.00 -
5.00\,GHz (Ji\v{r}i\v{c}ka et al. 1993). Figure 1 is the
spectrogram of the microwave type III pair train at frequency of
0.80 - 2.00 GHz during 06:21:30 - 06:22:00\,UT on 2011 September
26.

\begin{figure}       
\begin{center}
 \includegraphics[width=10.0 cm]{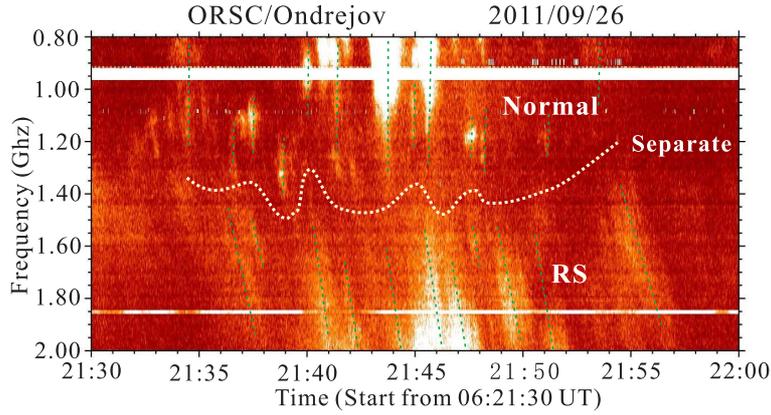}
 \caption{The spectrogram of microwave type III pair
train observed by ORSC/Ond\v{r}ejov on 2011 September 26. Green
dashed lines show the ridges of type III bursts. The white dashed
curve shows the separate frequency between the normal type III
branches and the RS type III branches.}
\end{center}
\end{figure}

The type III pair train is composed of a group normal type III
bursts and a group of RS type III bursts. The normal branches
start from about 1.30\,GHz (start frequency, $f_{st}$, defined as
the frequency at the start point of the microwave type III burst
with emission intensity exceeding the background significantly,
$>2\sigma$) and extends below 0.80 GHz with frequency drift
defined as the slope of the microwave type III burst on the
spectrograms, $D=\frac{df}{dt}$ ($f$ is the emission frequency) in
a range of 2.12 - 7.38\,GHz\,s$^{-1}$ (the relative frequency
drift $\bar{D}_{n}=\frac{df}{fdt}\sim$ 1.52 - 5.06 s$^{-1}$). The
RS branches start from about 1.50 GHz and extends beyond 2.00\,GHz
with frequency drift in a range of 281 - 647\,MHz\,s$^{-1}$
($\bar{D}_{r}\sim$ 0.23 - 0.50\,s$^{-1}$). That is to say, the
normal branches drift faster than the RS branches by about one
order of magnitude. The time difference between the start and end
of each type III burst is defined as the burst lifetime ($\tau$).
Here, the burst lifetime ranges from 0.14\,s to 1.10\,s with
average of 0.43\,s. The whole type III pair train lasts for about
20\,s.

From the observation we can see that there is a boundary frequency
between the normal and RS type III branches which is defined
separate frequency ($f_{x}$). For simple type III pair, $f_{x}$
can be obtained by reverse extending the normal and RS branches
and getting a crossing point. The frequency at the crossing point
is the separate frequency. For type III pair train, as there is no
clear one-to-one relationship between normal and the RS type III
branches, $f_{x}$ can be determined from the frequency at the
watershed line between the normal and RS type III group. Usually,
$f_{x}$ is a variable of time. We define $\frac{df_{x}}{dt}$ as
temporal change of separate frequency. The white dotted curve in
Figure 1 tracks the variation of the separate frequency ($f_{x}$)
during the type III pair train. $f_{x}$ is in a range of 1.225 -
1.494 GHz and has a lumpy variation. There is a frequency gap
($\delta f$) between the start frequencies of normal and RS
branches, which is in a range of 172 - 442\,MHz with average of
277\,MHz. Here, the separate frequency is nearly at the middle of
the frequency gap (shows in Figure 1).

ORSC observation has no circular polarization record. In the work
of Tan et al. (2016b), we made a comparison between the known- and
unknown-polarization events and deduced that the emission of the
above type III pair train is possibly weak polarization. It is
possibly the second harmonic plasma emission (the harmonic number
is $s=2$).

\subsection{Hard X-ray Observation}

There is no GOES soft X-ray (SXR) observation data during 05:15 -
06:28 UT around the type III pair train. We do not know what SXR
flare class of the burst. However, there is a M4.0 flare (start at
05:06\,UT, end at 05:13\,UT) before and a C4.7 flare (start at
07:28 UT, end at 07:38\,UT) after the above type III pair train in
active region NOAA 11302 (located at N12E22). At the same time,
there are two space telescopes observed a strong HXR burst around
the above microwave type III pair train.

One space telescope is the Ramaty High-Energy Solar Spectroscopic
Imager (RHESSI). The RHESSI data is processed with the software
developed by the RHESSI team (Lin et al. 2002) which have the
full-disk flux light curves and the imaging observations. This may
help us to determine the temporal evolutionary process of the
solar HXR burst associated to the microwave type III pair train
and its possible location on the solar disk. Panel (a) of Figure 2
presents the RHESSI HXR light curves in four energy bands (3 - 6,
6 - 12, 12 - 25, 25 - 50 keV). There is a strong HXR burst
starting at about 06:18\,UT, peaking at about 06:24\,UT and ending
at about 06:32\,UT. The low energy X-ray emission of 3 - 6\,keV is
relatively smooth with a maximum at about 06:25 UT (black curve in
the upper panel). The emission of 6 - 12\,keV has some small
spikes and reaches to a maximum at about 06:24 UT (red curve in
the upper panel). The high energy HXR emission of 12 - 25\,keV and
25 - 50\,keV have several sharp spikes during 06:21 - 06:24 UT,
and one of them occurred during 06:21:30 - 06:22:00 UT (the bottom
panel), just covered the above microwave type III pair train
(between the two vertical dotted lines).

\begin{figure}       
\begin{center}
 \includegraphics[width=6.0 cm]{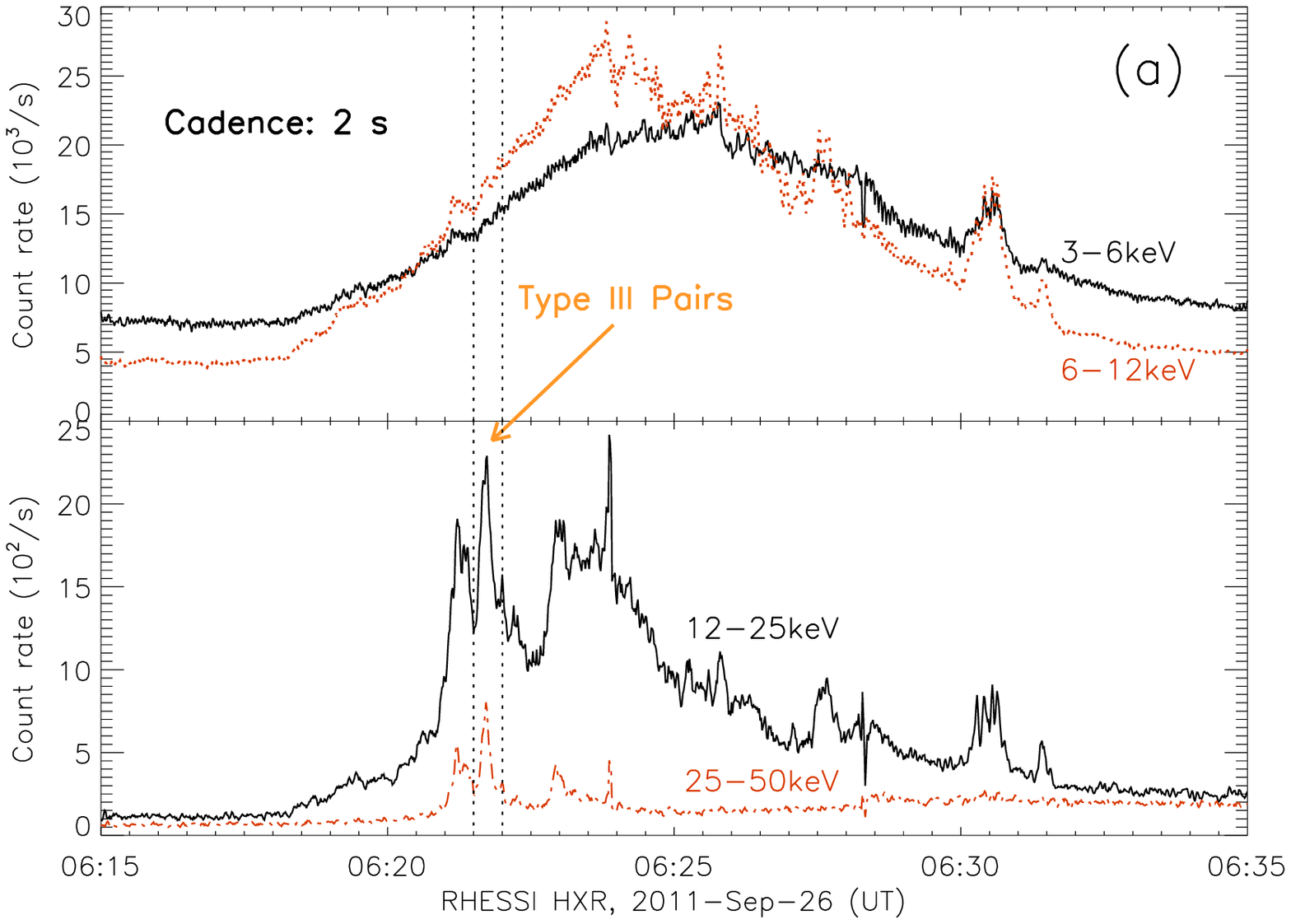}
 \includegraphics[width=6.0 cm]{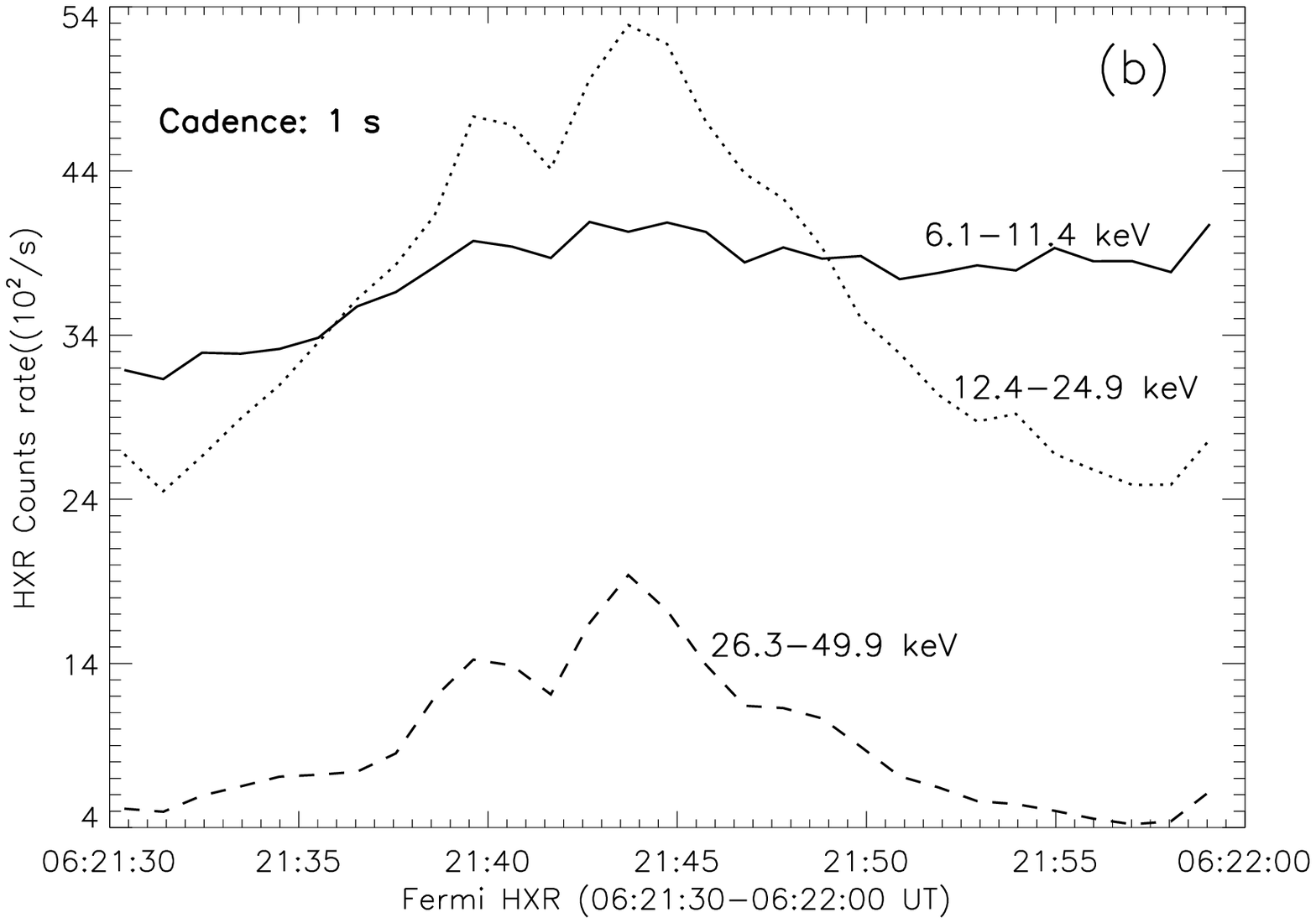}
 \caption{Panel (a): RHESSI HXR light curves in the four energy bands of 3 - 6, 6 - 12, 12 - 25 and 25 - 50 keV in the solar burst on 2011 September 26.
 The two vertical dotted lines shows the period of microwave type III pair train. Panel (b): Fermi HXR light curves in the energy bands of 6.1 - 11.4, 12.4 - 24.9,
 and 26.3 - 49.9 keV during the microwave type III pair train.}
\end{center}
\end{figure}

The other space telescope is the Fermi Gamma-Ray Burst Monitor
(GBM, Meegan et al., 2009). GBM is one of the instruments onboard
the Fermi Space Telescope (Atwood et al. 2009) launched on June
11, 2008. GBM is specifically designed for observing the whole
unocculted sky with a total of 14 scintillation detectors covering
the energy range from 8\,keV to 40\,MeV (Meegan et al. 2009). GBM
offers superb capabilities for the analysis of not only
$\gamma$-ray burst but solar flare HXR burst as well with cadence
of about 1 s. Panel (b) of Figure 2 shows the Fermi HXR light
curves in the energy bands of 6.1 - 11.4, 12.4 - 24.9, and 26.3 -
49.9 keV, respectively during the above microwave type III pair
train. Here, it shows that the microwave type III pairs are
closely related to the enhancements of nonthermal HXR emission at
energy of 12.4 - 24.9 and 26.3 - 49.9 keV. The low energy X-ray
emission (6.1 - 11.4 keV in panel b, and even 3 - 6 keV and 6 - 12
keV in panel a) has poor correlation to the microwave type III
pairs.

The radio type III pair train took place just at the impulsive
rising phase of the X-ray emission at energy of 3 - 12\,keV and
coincided with a sharp HXR spike at non-thermal HXR at energy of
above 12\,keV. These facts imply that the microwave type III pair
train is possibly produced by the same population of nonthermal
electrons which generated the HXR burst.

\subsection{Imaging Observation of the Source Region}

There are several telescopes having the imaging observations which
may provide the information of the location and structure of the
source region of the above HXR burst.

The left panel of Figure 3 presents the contours of HXR intensity
observed by RHESSI at energy of 6 - 12 keV (black), 12 - 25 keV
(green), and 25 - 50 keV (blue) integrated during 06:18:34 -
06:23:56 UT are overplotted on the magnetogram observed by the
Helioseismic and Magnetic Imager onboard the Solar Dynamic
Observatory (HMI/SDO) (Scherrer, et al. 2012) at 06:02 UT. HMI
observes the vector magnetic field of the full solar disk at
6173\AA~ with a spatial resolution of about 1$^{''}$ and cadence
from 90\,s to 135\,s. Although here AIA/SDO has no imaging
observation during 06:05 - 06:28 UT, we may use the above HMI
magnetogram to show the main magnetic properties of the source
region during the microwave type III pair trains, because
generally the magnetic polarity has no obviously change in a time
difference of about 20 min. The RHESSI HXR images was
reconstructed by using CLEAN algorithm (Hurford et al. 2002,
Schwartz et al. 2002). Here we can see that two SXR maxima of 6 -
12 keV overlayed on two regions with same magnetic polarity while
two HXR maxima of energy of 12 - 25 keV and 25 - 50 keV are
overlaying above the regions with opposite magnetic polarities
($P_{1}$ and $P_{2}$). It is possible that the HXR maxima of the
high energy 25 - 50 keV ($P_{1}$ and $P_{2}$) are very close to
the footpoints of the flaring loop. Their locations are
(-433$^{''}$, 172$^{''}$) and (-453$^{''}$, 114$^{''}$),
respectively. The distance between $P_{1}$ and $P_{2}$ is about
61$^{''}$, equals to about 4.4$\times10^{4}$ km. If we assume the
flaring loop is a semicircle and $P_{1}P_{2}$ is the diameter,
then the radius is approximated $R_{c}\sim$ 2.2$\times10^{4}$ km.

\begin{figure}        
\begin{center}
 \includegraphics[width=6.0 cm]{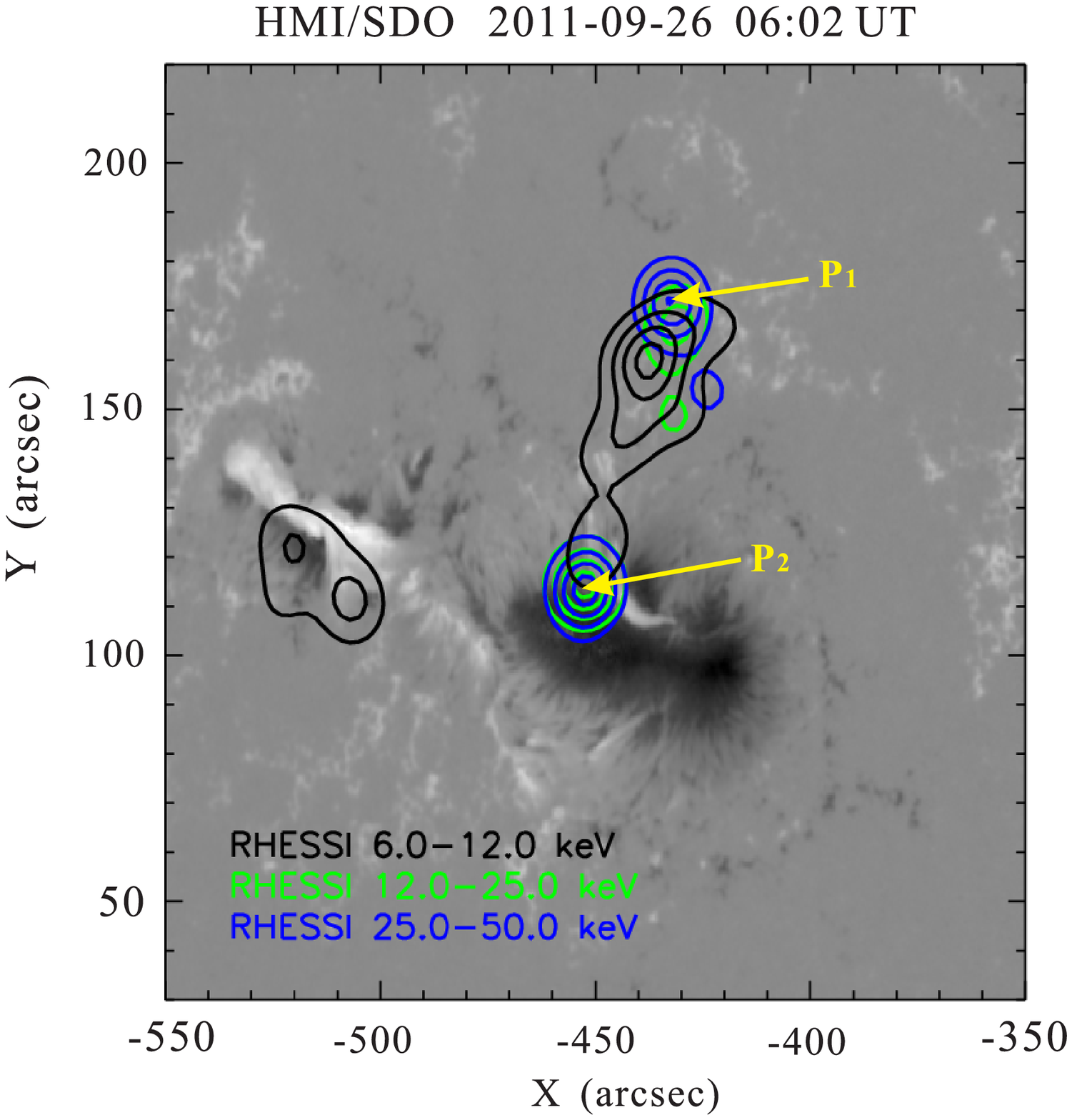}
 \includegraphics[width=6.0 cm]{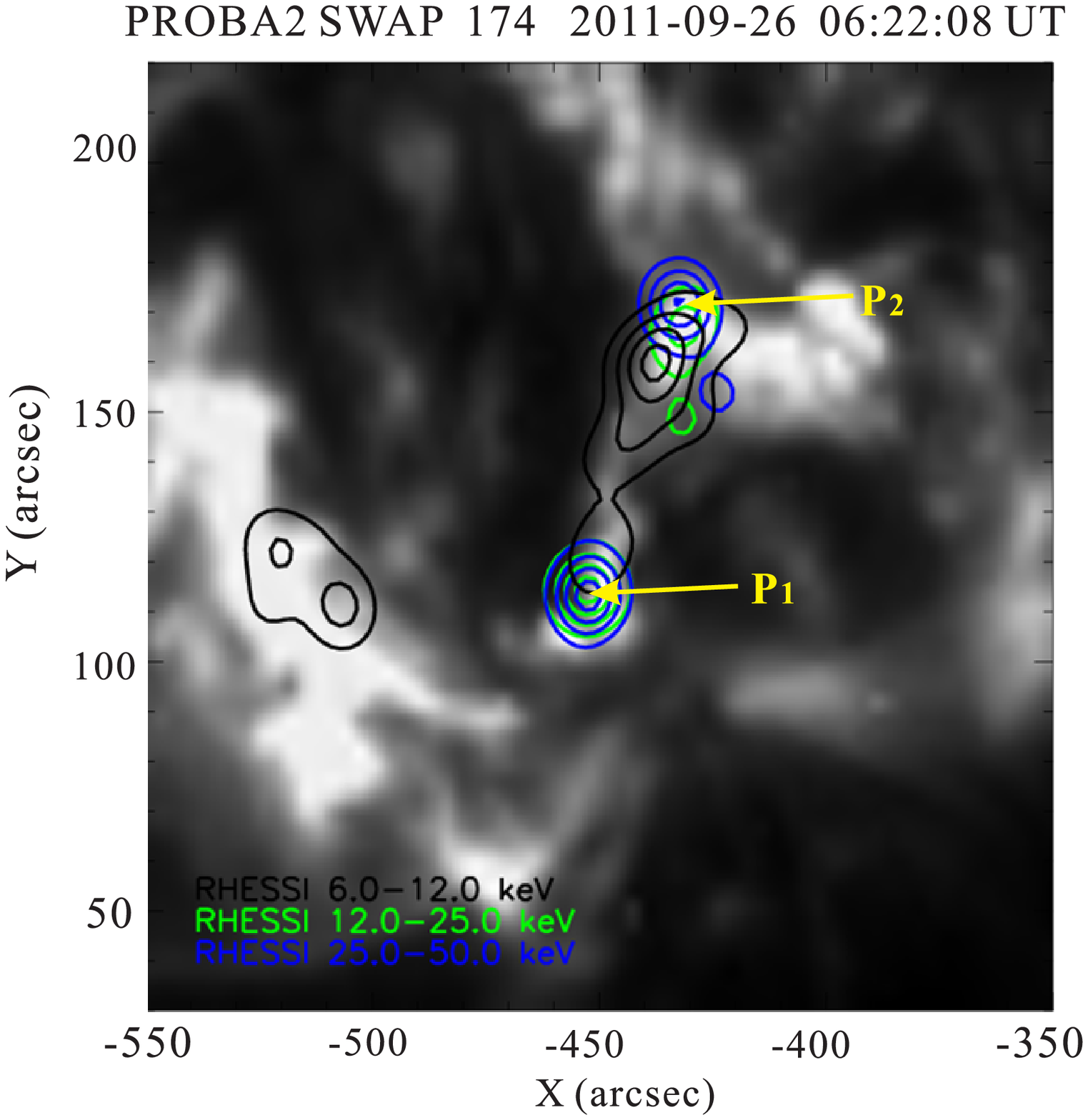}
 \caption{Source region of the solar burst. The background of the left panel is magnetogram observed by HMI/SDO at 06:02 UT
 while the background of the right panel is a 174 \AA~ EUV image observed by SWAP onboard PROBA-2 at 06:22:08 UT, 2011 September 26.
 The contours are HXR intensity at energy range of 6 - 12 keV (black), 12 - 25 keV (green), and 25 - 50 keV (blue) observed by RHESSI around the period of
 the microwave type III pair train. $P_{1}$ and $P_{2}$ are the footpoint sources with maximum intensities at 25 - 50 keV, respectively.}
\end{center}
\end{figure}

The background of the right panel in Figure 3 is a 174 \AA~ EUV
image observed by a compact solar EUV imaging telescope of the Sun
Watcher using APS detectors and image Processing (SWAP) instrument
onboard the Project for Onboard Autonomy 2 (PROBA2) satellite at
06:22:08 UT. PROBA-2 is a follow-on ESA micro-satellite technology
demonstration mission to the PROBA mission launched on 2011
October 21, and SWAP can obtain one image every minute (Berghmans
et al., 2006). The series of SWAP EUV images show a small
enhancement near the place of $P_{2}$ during the microwave type
III pair train. This fact might imply that the downward energetic
electron beams associated to the RS type III bursts may contribute
to heating the underlying plasmas near the footpoints of the
flaring loop.

\section{Diagnostics of the Source Region}
      \label{S-features}

Now with the above observation evidence we can obtain the physical
conditions near the source region of the above solar burst.

\subsection{Model and Method}

Recently, Tan et al. (2016a) proposed a new set of formulas to
diagnose the plasma density ($n_{e}$) and magnetic field $B$ near
the emission start sites of type III bursts and the velocity of
energetic electrons $v_{b}$ from the observations. As the
microwave type III pair train is composed of two branches of
microwave type III bursts, it is reasonable to apply them to
diagnose the physical conditions near the source region of primary
energy releasing and electron accelerations. Here we modified the
schematic diagram of the background conditions of microwave type
III pairs in Figure 4. The inflow (the big blue dotted arrows)
from the background plasmas (with magnetic field $B_{0}$) trigger
the magnetic reconnection and particle acceleration in the cusp
configuration or in the current sheet above the flaring loops
(around $C$). The upgoing and downgoing energetic electrons (the
solid red arrows) produce the plasma emission when they interact
with the background plasma. $U$, $D_{1}$ and $D_{2}$ are the start
sites of the upgoing and downgoing energetic electron beams,
respectively. The region between these start sites can be regarded
as source region of the solar burst (in the red dotted circle)
where magnetic reconnection occurs, electrons can be accelerated
to higher energy and the primary energy of the solar burst is
released. The distance from the acceleration site ($C$) to the
emission start site of the microwave type III bursts ($U$, $D_{1}$
and $D_{2}$) can be regarded as the acceleration length ($L_{c}$),
it is also the scale size of the source region. Under these
assumptions, then plasma density and magnetic fields near the
start sites are expressed (Tan et al. 2016a),

\begin{equation}
n_{e}=f_{st}^{2}/81s^{2} ~(m^{-3}),
\end{equation}

\begin{equation}
B_{L}<B<B_{H}
\end{equation}
$B_{L}=3.402\times10^{-19}(n_{e}T\bar{D}R_{c})^{\frac{1}{2}}$ and
$B_{H}=3.293\times10^{-16}[\frac{n_{e}T\bar{D}R_{c}}{(n_{e}\tau)^{\frac{1}{3}}}]^{\frac{1}{2}}$
are the lower and upper limits of the magnetic field,
respectively. $s$ is the harmonic number, when $s=1$ is the
fundamental plasma emission (always in strong polarization) and
$s=2$ is the second harmonic plasma emission (always in weak
polarization). $R_{c}$ is the curvature radius expressing the
divergence of magnetic field lines, $T_{e}$ is the plasma electron
temperature. The average of $B_{L}$ and $B_{H}$ can be regarded as
the best estimator of magnetic field:
$B\sim\frac{1}{2}(B_{L}+B_{H})$.

\begin{figure}      
\begin{center}
 \includegraphics[width=10.0 cm]{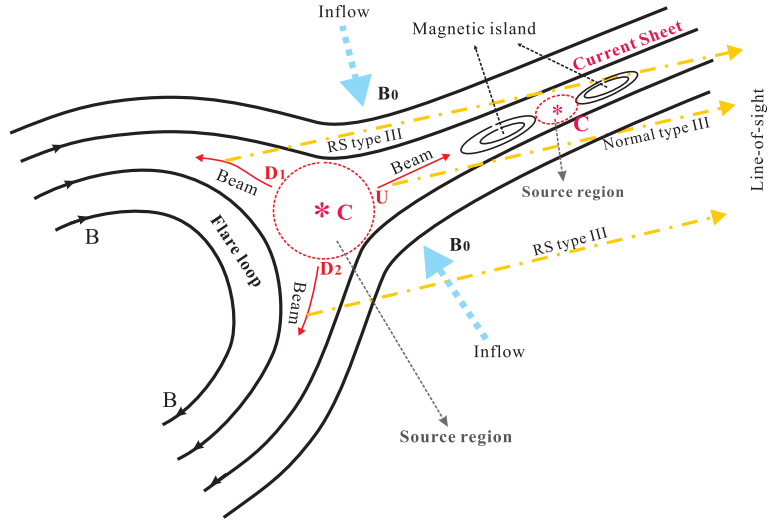}
 \caption{Schematic diagram of the background conditions of microwave type III burst
 pairs. The red arrows represent the energetic electron beams, the
 dash-dotted arrows represent the emission propagation. The red $D_{1}$, $D_{2}$ and $U$ are
 the start sites of the RS and normal type III bursts. The red star ($C$) is the magnetic reconnecting site.
 The blue arrows reflect the inflows which trigger the magnetic reconnections. The current sheet is located above the flare loop.}
\end{center}
\end{figure}

The velocity of the energetic electrons ($v_{b}$) can be
approximated (Tan et al. 2016a),

\begin{equation}
v_{b}\approx\frac{2\mu_{0}n_{e}k_{B}T}{B^{2}}\bar{D}R_{c},
\end{equation}
Here, $k_{B}$ is the Boltzmann constant. Furthermore, the energy
of the nonthermal electrons ($E$ in keV) can be derived
$E\approx256\frac{(v_{b}/c)^{2}}{\sqrt{1-(v_{b}/c)^{2}}}$, $c$ is
the speed of light.

Equation (3) indicates that the beam velocity is dominated not
only by the relative frequency drift rate, but also by the
magnetic field strength near the emission source region. Here, the
velocity is proportional to the relative frequency drift rate, and
inverse proportional to the square of magnetic field. This is
greatly different from the static barometric model which density
scale length only depends on the temperature of the solar medium
(Benz et al. 1983). It is also different from the isothermal
barometric model which is considering the free-free absorption and
the density scale length depends on both of temperature and
emission frequency (Dulk 1985, Stahli \& Benz 1987).

With the above results, the plasma $\beta$ ($\beta_{n}$ and
$\beta_{r}$) and density scale length ($H_{n}$ and $H_{r}$) near
the start sites can be estimated as
$\beta=\frac{n_{st}k_{B}T}{B^{2}/(2\mu_{0})}$ and
$H\approx\frac{1}{2}\beta\cdot R_{c}$, respectively. Furthermore,
the distance between the acceleration and start emission site of
the type III bursts can be estimated by,

\begin{equation}
L_{c}\approx H\cdot\frac{\delta f}{2f_{st}}.
\end{equation}

As the acceleration distance is related to the frequency
difference between the start frequency ($f_{st}$) and the separate
frequency ($f_{x}$), which is about a half of the frequency gap
($\delta f$). Therefore, there is a factor 2 in the denominator of
Equation (4).

As for type III pairs, we can obtain two sets of values associated
to the normal and RS branches ($f_{stn}$, $\bar{D_{n}}$, $n_{en}$,
$B_{n}$, $H_{n}$, $\beta_{n}$, $v_{bn}$, $E_{n}$, $L_{cn}$) and
($f_{str}$, $\bar{D_{r}}$, $n_{er}$, $B_{r}$, $H_{r}$,
$\beta_{r}$, $v_{br}$, $E_{r}$, $L_{cr}$). Here, the subscripts
$n$ and $r$ are indicated the normal and RS type III branches,
respectively. $v_{bn}$ and $v_{br}$ are the velocities of the
upgoing and downgoing electron beams while $E_{n}$ and $E_{r}$ are
their energies, respectively.

Here, it is worth to note that there are two kinds of reconnection
sites, one is located in the cusp configuration near the loop top,
the other is in the current sheet above the flare loops. As the
current sheet may trigger the tearing-mode instability and produce
the quasi-periodic pulsating structures in the microwave bursts
(Kliem et al. 2000). In this work the type III pair train has no
such pulsating property, therefore we tend to suppose the
reconnection site is possibly located in the cusp configuration.

\subsection{Diagnostic Results}

Using Equation (1) - (4), we may estimate the physical conditions
near source region of the solar bursts. Here, the plasma
temperature is derived from the X-ray emission observed by
GBM/Fermi with cadence of about 1 s (Meegan et al., 2009). The
background is subtracted as a values of non-flare periods. The
model consists of thermal function (the optically thin thermal
bremsstrahlung radiation function for one temperature) and a thick
target model in power-law function. The derived temperatures are
in a range of 12 - 24 MK during the microwave type III pair train
(the black solid curve in panel (a) of Figure 5).

\begin{figure}     
\begin{center}
 \includegraphics[width=10 cm]{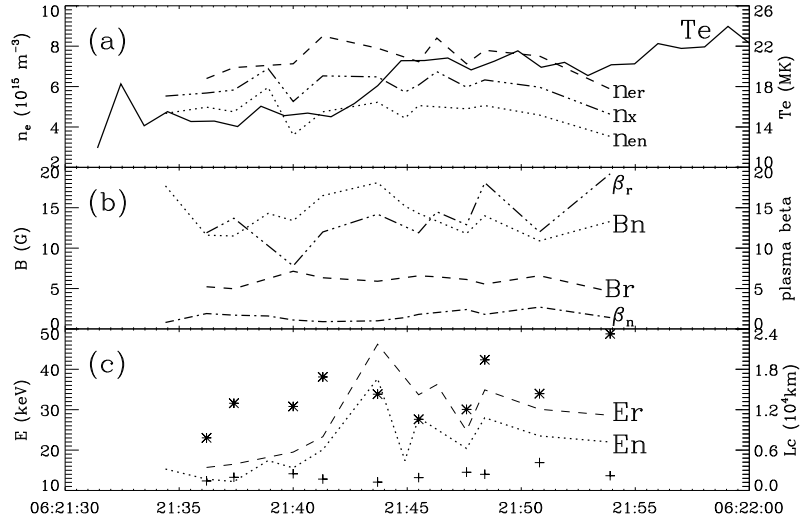}
 \caption{Evolution of the physical conditions around the microwave type III pair train. $n_{en}$, $n_{er}$ are the plasma density near
 the start sites of the normal and RS type III branches, respectively. $n_{x}$ is the plasma density near the acceleration site.
 $T_{e}$ is the plasma temperature. $B_{n}$ and $B_{r}$ are magnetic
 field, $\beta_{n}$ and $\beta_{r}$ are the plasma beta value near start sites of normal and RS type III branches, respectively.
 $E_{n}$ and $E_{r}$ are the energy of the upgoing and downgoing electrons, respectively. The plus ($+$) and star signs ($\ast$) in panel (c) are the
 acceleration lengths corresponding to the upgoing and downgoing electrons, respectively.}
\end{center}
\end{figure}

The plasma densities near the emission start sites of the normal
and RS type III bursts are over-plotted in dotted and dashed
curves in panel (a) of Figure 5, respectively. The density near
the start sites of the normal type III bursts is in a range of
(3.5 - 6.0)$\times10^{15} m^{-3}$, and (5.9 - 8.5)$\times10^{15}
m^{-3}$ near the start site of RS type III bursts. We may derive
the plasma density near the electron acceleration site ($n_{x}$)
which is the middle between the above two values, that is around
6$\times10^{15} m^{-3}$ with little variations (associated to the
separate frequency $f_{x}$ showed by the white solid curve).

The derived magnetic fields near the emission start sites of the
type III bursts are showed in panel (b) of Figure 5. It is
interesting that the magnetic field near the start site of the RS
type III branches is in a range of 4 - 8 Gauss, weaker than that
near the start site of the normal type III branches (9 - 18
Gauss). As we know that RS type III bursts are related to the
downgoing electron beams which start site may locate below the
acceleration site while the start site of the upward electron beam
associated to normal type III bursts is above the acceleration
site. Usually, the coronal magnetic field is decreasing with the
height which is conflicting with the above results. We may explain
this conflict as following: in magnetic reconnection regime, the
anti-paralleled component of magnetic field near the reconnection
site (C) is very close to 0 where is a magnetic singular point and
electron acceleration takes place around it (the other place is in
the center of current sheet where the magnetic field tends to be
near 0). The magnetic field increases from the reconnection site
to the emission start sites of microwave type III burst ($U$,
$D_{1}$ and $D_{2}$), which is obviously different from the
general coronal magnetic field.

The velocities of the upgoing energetic electrons are in a range
of 0.22 - 0.37\,c, a bit smaller than that of the downgoing
energetic electrons (0.24 - 0.41\,c). The corresponding energies
are 12 - 38\,keV and 16 - 47\,keV for the upward and downward
electron beams, respectively (panel (c) of Figure 4). Here, we
find that although the RS type III branches drift much slower at
about one order of magnitude than the normal type III branches,
but the corresponding energy difference of the downgoing and
upgoing electrons is only from 2.7 keV to 9.5 keV, which is
relatively very small.

Many previous publications show that the energies in both upgoing
and downgoing electrons are similar to each other. There is no
remarkable difference between the normal and RS type III branches.
This is mainly because their relative frequency drift rates of the
normal and RS type III branches are similar to each other in their
observations (Dulk 1985, Benz et al. 1992, Aschwanden et al. 1995,
Meshalkina et al. 2004, etc.). However, in the flare event on
2011-09-26, we found that there is an order of magnitude apart of
the relative frequency drift rates between the normal and RS
branches. With such great different drift rates, we still obtained
similar energies of the upgoing and downgoing electrons. The
reason is that although the relative drift rates between the two
branches have an order of magnitude apart, but their magnetic
field strengths also have several times apart, and these make it
is possible to yield similar beam velocities. This can be
explained by using Equation (3).

The little difference between the upgoing and downgoing electrons
indicates that they may accelerate possibly by the similar
mechanism and more effective to the downgoing electrons in the
magnetic reconnection region. In the work of Tan et al. (2016b),
the same method was applied to the microwave type III pair trains
in the postflare phase of an X3.4 flare on 2006 December 13 and
found that both of the upgoing and downgoing electrons have almost
same energy of 42 - 64 keV.

The small difference of the acceleration in the above two flare
events is possibly because the two microwave type III pair events
take place in different phase in the solar flares, one is occurred
in an impulsive rising phase and the other is in a postflare decay
phase. In this work, microwave type III pair train occurs in the
impulsive rising phase of the solar HXR burst where the magnetized
plasma is possibly much more instable and the magnetic
reconnection is stronger. This point can be implied indirectly
from the plasma beta ($\beta_{p}$). In this work $\beta_{p}$ are
in a ranges of 0.8 - 2.6 and 7.5 - 18.5 in the start sites of the
upgoing and downgoing electron beams, respectively. The large
value of $\beta_{p}$ implies the more violent instability in the
magnetized plasma (Drake et al. 2010, Schoeffler et al. 2011). The
particle acceleration may be more effective to the downgoing
electrons than to the upgoing electrons in the flare impulsive
phase. In the X3.4 flare of 2006 December 13, the microwave type
III pair trains took place in the postflare decay phase and the
$\beta_{p}$ in the start site of upgoing electron beams (0.87 -
1.05) was very close to that of the downgoing electron beams (1.09
- 2.74). The little difference of $\beta_{p}$ indicates that their
magnetic reconnection are at the similar level. Therefore the
particle acceleration is also similar to each other.

By using equation (4) we make an estimation of the scale size of
acceleration region ($L_{c}$) which is in a range of (1.2 -
4.1)$\times10^{3}$ km (the plus signs $+$ in panel (c) of Figure
5) for the upgoing electron beams and (0.8 - 2.3)$\times10^{4}$ km
for the downgoing electron beams (the star sigh $\ast$ in panel
(c) of Figure 5). This means that type III bursts begin to
generate just after the energetic electron beam propagating a
considerable distance from the acceleration site.

\section{Summary and Discussions}
      \label{S-features}

Using the observations of solar radio spectrometers, hard X-ray
emission of RHESSI and Fermi and EUV images of PROBA 2, we
diagnosed the physical parameters near the start site of normal
and RS type III bursts, which includes the plasma density,
temperature, magnetic field, and plasma beta near the start sites
of the microwave type III burst, and the primary energy of the
upgoing and downgoing electrons and the distance between the
acceleration and emission start sites of the type III bursts,
respectively. This is the first time to derive so much abundant
information of the source region of solar burst directly from the
observations. These results present several properties:

(1) The plasma density associated to the start site of RS type III
branches is higher than that of normal type III branches, while
the magnetic field associated to the emission start site of RS
type III branches is weaker than that of normal branches, and
these lead to the plasma beta values near the start site of RS
type III branches higher than that of the normal branches.

(2) Although the RS type III branches drift slower at about one
order of magnitude than the normal type III branches, the energies
of the downward electron beams are still very close to that of the
upward electron beams.

(3) The plasma density, temperature, magnetic field strength, and
the distance between the acceleration and the emission start sites
almost have no obvious variations during the period of type III
pair trains, while the derived energy of electrons has an obvious
peak value which is consistent to the hard X-ray emission during
the half minute of type III pair burst.

These facts imply that both of the upgoing and downgoing electron
beams are accelerated by similar mechanism in the magnetic
reconnection region, and their little difference is just because
of their different background conditions, including the plasma
density, magnetic field strength and the scale size of the source
region, etc.

The above calculations indicate that most of the plasma beta near
the emission start sites of microwave type III pair bursts are
around or even greatly higher than an unity. This fact may
reflects the highly-dynamic properties of the source region where
magnetic reconnection and particle accelerations (by the
reconnecting electric field) take place. Actually, the magnetic
field tend to be very weak and the plasma beta may become very
large near the reconnecting site. Around these sites, not only the
magnetic reconnection and electron acceleration take place, but
also the plasma turbulence and various plasma instabilities occur.
Therefore it is much different and complicated from the plasma in
the coronal background and flaring loops.

Actually, Equation (2) indicates $B\propto R_{c}^{1/2}$, and
plasma beta value $\beta_{p}\propto \frac{1}{R_{c}}$. Figure 4
shows that the upward electron beams may meet the magnetic islands
in the current sheet, and here the magnetic curvature radius
$R_{cn}$ will possibly become smaller than the radius of the flare
loop. We adopt the flare loop to approximate $R_{c}$ in both
normal and RS type III burst regimes, this will overestimate
$R_{c}$ and the magnetic field near the start site of the upward
beams, and underestimate the plasma beta. Equation (3) indicates
$v_{b}\propto R_{c}B^{-2}$ which may derive that the velocity
($v_{b}$) is independent to the magnetic field strength. So, the
uncertainties of magnetic field will not affect the estimation of
the velocity and energy of the electrons.

There are still uncertainties in estimating the magnetic field
near the emission start site of the normal type III branches.
Equation (2) indicates that magnetic field estimation ($B$)
depends on the curvature radius $R_{c}$ of the magnetic field
lines. $R_{c}$ for the normal type III branches should be
different from that for RS type III branches. Figure 4 shows that
the start sites ($U$) of the upward electron beam locate above the
magnetic reconnecting site ($C$) and the start sites ($D_{1}$ and
$D_{2}$) of downward electron beam locate below the magnetic
reconnecting site. $D_{1}$ and $D_{2}$ are very close to the top
of flare loop while $U$ is close to the current sheet above the
flare loop. In current sheet, the tearing-mode instabilities may
take place and produce magnetic islands and plasmoids which can
result in the complex magnetic structures (Kliem et al. 2000).
Therefore, the $R_{c}$ of the magnetic field lines near the site
of $U$ may be different to that near $D_{1}$ and $D_{2}$. However,
so far we have no reasonable method to determine exactly the
curvature radius of magnetic field lines near the site of $U$
where the upward electron beams pass through. As an expedient, we
simply use the flare loop radius to approximate it. In the near
future, it is possible that we can obtain the curvature radius
directly from the advanced imaging observation for the normal and
RS type III branches, for example the Mingantu Spectral
Radioheliograph (MUSER, the former name is CSRH, Yan et al. 2009).

\begin{acknowledgements}
The authors are grateful to the referee's kindly and helpful
comments for improving this paper. We also thank the ORSC teams in
Ond\v{r}ejov for providing observation data. This work is
supported by the NSFC Grants 11273030, 11373039, 11433006,
11573039, and 2014FY120300, CAS XDB09000000, the Grant
P209/12/00103 (GA CR) and Project RVO: 67985815 of the
Astronomical Institute AS, as well as by the Marie Curie
PIRSES-GA-295272-RADIOSUN project.

\end{acknowledgements}

\end{article}
\end{document}